\documentstyle[aps,prl,epsf,floats,amssymb,cite]{revtex}

\def\CM{{\cal M}}
\def\tr{{\rm Tr\, }}
\def\pf{{\rm Pf\, }}
\def\im{{\rm Im\, }}

\def\c{{\Bbb C}}
\def\z{{\Bbb Z}}

\def\ssqr#1#2{{\vbox{\hrule height #2pt
\hbox{\vrule width #2pt height#1pt \kern#1pt\vrule width #2pt}
\hrule height #2pt}\kern- #2pt}}

\def\sqr{\mathchoice\ssqr8{.4}\ssqr8{.4}\ssqr{5}{.3}\ssqr{4}{.3}}

\def\aa{\vbox{\hbox{$\sqr$}
\nointerlineskip\kern-.3pt\hbox{$\sqr$}}}

\begin{document}

\tightenlines

\twocolumn[\hsize\textwidth\columnwidth\hsize\csname@twocolumnfalse\endcsname

\preprint{\vbox{
\hbox{UCSD/PTH 98--19}
\hbox{hep-th/9805144}
}}

\title{Wess-Zumino Terms in Supersymmetric Gauge Theories}
\author{Aneesh V.~Manohar}
\address{Department of Physics, University of California at San Diego,\\
9500 Gilman Drive, La Jolla, CA 92093-0319}
\date{May 1998}
\maketitle
\begin{abstract}
The Wess-Zumino term is constructed for supersymmetric QCD with two colors and
flavors, and is shown to correctly reproduce the anomalous Ward identities.
Supersymmetric QCD is also shown not to have topologically stable skyrmion
solutions because of baryon flat directions, which allow them to unwind. The
generalization to other supersymmetric theories with quantum modified
constraints is discussed.
\end{abstract}
\pacs{PACS}

]\narrowtext

The configuration space of zero-energy states of a supersymmetric gauge theory
is known as the moduli space $\CM$, and is parameterized by the expectation
values of gauge invariant composite fields subject to constraints. If $\CM$ has
non-trivial topology, there can exist topological terms in the effective action
such as a Wess-Zumino term~\cite{wz,witten}. It is also possible to have
topologically stable field configurations, such as skyrmions or
vortices~\cite{skyrme,witten}. The early work on Wess-Zumino terms in
supersymmetric gauge theories~\cite{nr,ggs,mcarthur,harada} was done before the
recent work  of Seiberg and others elucidating the structure of the quantum
moduli spaces~\cite{seiberg}. The existence of topological terms is
re\"examined in light of these results. This paper concentrates on studying the
Wess-Zumino term in supersymmetric QCD with two colors and two flavors, the
simplest example of a theory with a quantum deformed moduli-space. The results
can be generalized to other quantum deformed theories~\cite{ben}.

The structure of the moduli space of supersymmetric QCD in $3+1$ dimensions
depends on the number of colors $N$ and flavors $F$. For $F<N$ the low-energy
description is in terms of the expectation value of gauge invariant mesons
$M^i{}_j$, and the effective theory has a non-perturbative superpotential
\[
W = (N-F) \left({\Lambda^{3N-F} \over \det M}\right)^{1/(N-F)},
\]
where $\Lambda$ is the non-perturbative scale parameter of the theory. The
quantum theory is unstable, with $\det M \to \infty$.  The moduli space for
$\det M \not =0$ is isomorphic to the group $GL(F,\c)$. This example has been
studied in detail in the literature~\cite{nr,ggs,mcarthur,harada}, and the
analysis will not be repeated here. The moduli space $\CM = GL(F,\c)$ has
a Wess-Zumino term, and supports stable skyrmion solutions.

The cases we will examine are $F=N$ and $F=N+1$, where the quantum moduli
spaces have recently been constructed~\cite{seiberg}. For $N=F+1$, the moduli
space~\cite{seiberg} is given by the expectation values of gauge invariant
mesons $M^i{}_j$, baryons $B_i$, and antibaryons $\tilde B^j$. Here $i,j=1,
\ldots,F$ are flavor indices. There are non-trivial constraints among the basic
invariants,
\begin{equation}\label{2}
B_i M^i{}_j=0,\ M^i{}_j\tilde B^j=0,\ {\rm cof}{}_i{}^j M = B_i \tilde B^j,
\end{equation}
where ${\rm cof}{}_i{}^j M$ is the cofactor of the $ij$ entry of $M$. These
constraints are precisely the same as those obtained by minimizing the
superpotential
\begin{equation}\label{3}
W = {B_i M^i{}_j \tilde B^j - \det M \over \Lambda^{2N-1}}.
\end{equation}
This theory clearly has a topologically trivial moduli space. One can make a
deformation retract of the moduli space to the origin $M=B = \tilde B=0$ since
all the constraints are homogeneous. It is therefore not possible to construct
a topological term in the effective action.  Obviously, a similar result holds
for any theory whose moduli space is given by gauge invariants subject to
homogeneous constraints, such as s-confining theories~\cite{sconf},  or any
theory whose moduli space has no constraints, such as those with an affine
moduli space~\cite{affine}. In these theories, the flavor anomalies of the
gauge-invariant composites agree with those of the microscopic fields, so a
Wess-Zumino term is not required in the low-energy theory.

The interesting case is supersymmetric QCD with $F=N$; the $F=N=2$ case will be
studied here. Since the ${\bf 2}$ and ${\bf \bar 2}$ representations of $SU(2)$
are equivalent, the quarks and antiquarks can be combined to form four $SU(2)$
doublets. The flavor symmetry of the theory is $SU(4) \times U(1)_R$. The
mesons and baryons can be combined into a single $4\times 4$ antisymmetric
matrix $V$,
\begin{equation}
V=\left(\begin{array}{c|c}
\begin{array}{cc}
0 & B \\
-B & 0 \\
\end{array} & M \\ \hline
- M^T & \begin{array}{cc}
0 & \tilde B \\
-\tilde B & 0 \\
\end{array}
\end{array}\right)
\end{equation}
which transforms as the two-index antisymmetric tensor under flavor-$SU(4)$,
and has zero $R$-charge. The quantum constraint is~\cite{seiberg}
\begin{equation}\label{6}
\pf V= \Lambda^4,
\end{equation}
where ${\rm Pf}$ is the Pfaffian. The constraint can be written as
\begin{equation}
B \tilde B -\det M = B \tilde B - M_{11} M_{22} +  M_{12} M_{21} =\Lambda^4.
\label{7}
\end{equation}
It is straightforward to determine the topology of the quantum moduli space
$\CM$ given by $V$ subject to the constraint Eq.~(\ref{6}). The $SU(4)$ group
is equivalent to $SO(6)$, and $V$ is the ${\bf 6}$ dimensional (i.e. vector)
representation of $SO(6)$, which can be denoted by $\left(X_1,\ldots
,X_6\right)$, where the $X_i$ are linear combinations of the $V_{ij}$. The
constraint Eq.~(\ref{6}) is the $SO(6)$ invariant constraint
\begin{equation}\label{16}
\sum_{i=1}^6 X_i^2 = \Lambda^4,
\end{equation}
and the moduli space $\CM$ is the surface in $\c^6$ given by Eq.~(\ref{16}). It
is straightforward to show that there is a deformation retract of $\CM$ onto
the real section given by taking $X_i$ real, i.e. the five-sphere $S^5$.

The homotopy and cohomology groups of $\CM$ are identical to those of $S^5$. In
particular $H^5(\CM)=\z$ and $\pi_3(\CM)=0$, so that one can write down a
Wess-Zumino term, but there are no topologically stable skyrmion solutions. By
analogy with QCD (which has stable skyrmion solutions), one can write down a
``skyrmion'' field configuration $V({\bf x})$ which is a static field
configuration from $S^3 \to \CM$,
\begin{equation}\label{40}
B=\tilde B =0, \qquad M = \exp\left[ i {\boldmath \tau} \cdot {\bf {\hat x }} F(\left|
{\bf x} \right|)\right],
\end{equation}
with $F(\infty)=0$, $F(0)=\pi$. The skyrmion has non-trivial winding number if
one only looks at the subspace $B =\tilde B =0$, $\det M\not= 0$, but can
unwind because of the baryon directions $B$ and $\tilde B$. It is easy to
explicitly write down a sequence of field configurations that go through the
point $B \tilde B=1$, $\det M=0$, and allows the skyrmion to unwind. This
result generalizes to $F=N>2$, where the moduli space is given by a $N\times N$
matrix $M$, and baryons $B$ and $\tilde B$ with the quantum
constraint~\cite{seiberg}
\begin{equation}\label{30}
\det M - \tilde B B = \Lambda^{2N}.
\end{equation}
(The difference in sign from Eq.~(\ref{7}) has to do with the relation between
the ${\bf 2}$ and ${\bf \overline 2}$ representations of $SU(2)$, and in
unimportant.) A skyrmion Eq.~(\ref{40}) embedded in the first $2 \times 2$
block of $M$ can unwind because of the baryon directions.

A Wess-Zumino functional $\Gamma$ on $\CM$ can be defined following the method
used by Witten~\cite{witten} for QCD. A field configuration $V(x)$ is a map 
$V: S^4 \to \CM$ from spacetime to the moduli space. Since $\pi_4(\CM)=0$, the
four-surface in $\CM$ given by the image of space-time under $V$ is the
boundary of a five-surface $\Sigma_5$ in $\CM$. The Wess-Zumino functional is
given by integrating a closed (but not exact) five-form $\omega_5$ defined on
$\CM$ over the five-surface $\Sigma_5 \in \CM$. As for QCD, one finds that
since $\pi_5(\CM)=\Bbb{Z}$,  the Wess-Zumino action is ambiguous. The ambiguity
in $\Gamma$ is an integer times the integral of $\omega_5$ over the five-sphere
that generates $\pi_5(\CM)$. The ambiguity is irrelevant for the quantum theory
provided $\exp i\Gamma$ is well-defined. This determines the Wess-Zumino term
to be
\begin{equation}\label{wz}
\Gamma =  {1\over 240 \pi^2} \im \int_{\Sigma_5} \tr \left(V^{-1} dV \right)^5.
\end{equation}
The term in the effective action is $n \Gamma$, where $n$ is an integer. The
Wess-Zumino term is well-defined, since the constraint Eq.~(\ref{6}) implies
that $V$ is invertible. The coefficient is fixed by requiring that the integral
over the five-sphere is $2 \pi$. Equation~(\ref{wz}) gives the bosonic part of
the Wess-Zumino action; one can always make the action supersymmetric by adding
fermionic components, and writing the Wess-Zumino term as a
$D$-term~\cite{nr,ggs,mcarthur,harada}. In the remainder of the paper, we will
concentrate only on the bosonic part of the Wess-Zumino term, since that is the
piece relevant for the anomalous Ward identities. The Wess-Zumino term
Eq.~(\ref{wz}) has been written using a holomorphic five-form.  For a
discussion of why this is possible, see section~5 of Ref.~\cite{nr}.

The Wess-Zumino term has been written as the imaginary part of the integral in
Eq.~(\ref{wz}). Only the imaginary part has a quantized coefficient and
contributes to the anomalous Ward identities. The real part is an allowed term
in the effective action, and does not have a quantized coefficient since its
integral over $S^5$ vanishes. The real part vanishes in QCD, because the
Wess-Zumino action is written as an integral of the form Eq.~(\ref{wz}), with
$V \to U$, a unitary matrix. Here $V$ is not unitary, and the integral has both
real and imaginary parts.

The integer $n$ multiplying Eq.~(\ref{wz}) in the effective action is fixed by
requiring that the low-energy theory reproduce all the flavor anomalies of the
microscopic theory. (The microscopic theory refers to the theory written in
terms of quarks and gluons, and the low-energy theory refers to the theory
written in terms of gauge invariant mesons and baryons.) The microscopic theory
has a $SU(4) \times U(1)_R$ flavor symmetry.  The $U(1)_R$, $U(1)_R^3$ and
$SU(4)^2 U(1)_R$ anomalies match between the microscopic and low-energy
theories when computed using the massless fermions in the two theories, but the
$SU(4)^3$ anomalies do not match. This was the original motiviation for
introducing the quantum deformation Eq.~(\ref{6}) in the moduli space $\CM$.
The $SU(4)$ symmetry is broken at all points on $\CM$, so the $SU(4)^3$
anomalies computed using the massless fermions of the microscopic and
low-energy theories need not match. Nevertheless, the anomalous Ward identities
must be satisfied~\cite{thooft}. Anomalous Ward identities get contributions
from massless fermions and Goldstone bosons~\cite{coleman}, so the $SU(4)^3$
Ward identity gets an additional Goldstone boson contribution from the
Wess-Zumino term, which fixes $n$.

The contribution of the Wess-Zumino term can be determined by turning on weakly
coupled background gauge fields for the $SU(4) \times U(1)_R$ flavor symmetry,
and studying the variation of the Wess-Zumino term under local  flavor symmetry
transformations. The variation of $\Gamma$ under an infinitesimal local $SU(4)
\times U(1)_R$ transformation is
\begin{eqnarray}
\delta \Gamma &=& {1 \over 48 \pi^2} \int_{\partial \Sigma_5} \tr \left[
d \epsilon^T\left(V^{-1} dV \right)^3 - d \epsilon \left(dV V^{-1}\right)^3
\right] \nonumber \\
& =& -{1 \over 24 \pi^2} \int_{\partial \Sigma_5}\tr  d \epsilon \left(dV V^{-1}\right)^3,
\label{var}
\end{eqnarray}
where $\epsilon = \epsilon^a T^a$ is an $SU(4)$ generator,
so that the Wess-Zumino term contributes to the $SU(4)$ flavor current,
\begin{equation}
j_{\mu}^a = {1 \over 24 \pi^2} \epsilon_{\mu \nu \alpha \beta}
\tr T^a
\left(\partial^\nu V V^{-1}\right)\left(\partial^\alpha V V^{-1}\right)
\left(\partial^\beta V V^{-1}\right).
\label{curr}
\end{equation}

The field $V$ does not transform under $U(1)_R$, so that one might naively
think that the Wess-Zumino term does not contribute to the $R$ current.
However, for QCD, Witten pointed out that the Wess-Zumino term contributes to
the baryon number current even though it is written in terms of mesons which
have zero baryon number. This subtlety does not occur here.  The possible
Wess-Zumino contribution to the $R$ current is given by using 
Eq.~(\ref{curr}), and replacing $\epsilon^a T^a \rightarrow \epsilon \openone$,
as for QCD.  The resulting current vanishes, since
\begin{equation}
\tr \left(dV V^{-1} \right)^3=0,
\label{20}
\end{equation}
because $V$ is an antisymmetric matrix. To see this, note that
\begin{eqnarray*}
\tr \left(dV V^{-1} \right)^{2m+1} &=& \left(-1\right)^{m}
\tr \left(dV V^{-1} \right)^{T 2m+1} \\
&=& \left(-1\right)^{m}
\tr \left(dV V^{-1} \right)^{2m+1}
\end{eqnarray*}
since transposing the matrices changes the ordering of the differential forms.
This shows that $\tr \left(dV V^{-1} \right)^3$ vanishes, but not  $\tr
\left(dV V^{-1} \right)^5$. Thus the Wess-Zumino term does not contribute to
anomalous Ward identities involving $U(1)_R$, which is consistent with the fact
that the low-energy fermions correctly reproduce the anomalies of the
microscopic theory which involve $U(1)_R$. The vanishing of Eq.~(\ref{20}) is
related to $\pi_3(\CM)=0$ and the non-existence of skyrmions.

The Wess-Zumino term in the presence of background $SU(4)$ gauge fields $A$ can
be obtained from the result of Witten for QCD~\cite{witten} (in the parity
invariant form of Ref.~\cite{moore}). It is conventionally written as
\[
\Gamma(V) + {1\over 48 \pi^2} Z(V,A),
\]
where $Z$ can be obtained from the one in Ref.~\cite{moore} by the replacement
\begin{equation}\label{repl}
A_L \to A,\ A_R \to -A^T,\ \Sigma \to V,\ \Sigma^\dagger \to V^{-1}.
\end{equation}
The
variation under a gauge transformation is
\begin{eqnarray}
\delta\left(\Gamma + {1\over 48 \pi^2} Z\right) &=& 
-{1 \over 24 \pi^2} \int \tr \epsilon d\left(A dA + {1\over 2}  A^3 \right)
\nonumber \\
&&\hspace{-1.5cm}+ {1 \over 24 \pi^2} \int \tr -\epsilon^T d\left(A^T dA^T -
 {1\over 2}  A^{T3} \right) \label{anom1} \\
 &=& -2{1 \over 24 \pi^2} \int \tr \epsilon d\left(A dA + {1\over 2} 
  A^3 \right)\nonumber 
\end{eqnarray}
using the results of Ref.~\cite{witten} and Eq.~(\ref{repl}). Thus the
Wess-Zumino functional contribution to the $SU(4)^3$ anomaly is twice that of a
Weyl fermion in the fundamental representation of $SU(4)$. The quarks in the
microscopic theory are a ${\bf 4}$ of $SU(4)$, and contribute an anomaly of two
(since they are a gauge $SU(2)$ doublet). The low-energy field fermions are a
${\bf 6}$ of $SU(4)$, which is a real representation, and so do not contribute
to the anomaly. This determines the coefficient $n$ in front of the Wess-Zumino
term in the effective action to be $n=1$.  Note that $\Gamma$ contributes to
the anomalous Ward identity even though $V$ transforms as a real representation
of $SU(4)$, and that $n=1$ for two colors, unlike in QCD where $n=2$.

At points on the moduli space where $V=J$,
\[
J = \left( \begin{array}{cc}
i\sigma_2 & 0 \\ 0 & i \sigma_2 \\
\end{array} \right),
\]
the flavor $SU(4)$ group is broken to an $Sp(4)$ subgroup. The $Sp(4)^3$ flavor
anomaly matching condition is satisfied between the fermions in the high-energy
and low-energy theories. One can check that Eq.~(\ref{anom1}) does not
contribute to the anomalous Ward identity for $Sp(4)$. For the $Sp(4)$
subgroup, $J \epsilon J = \epsilon^T$, $J A J = A^T$. Using these relations, it
is easy to see that Eq.~(\ref{anom1}) vanishes. This must be the case, since
$Sp(4)$ does not have complex representations.

The above results are easily generalized to $Sp(2n-2)$ theories with $2n$
fundamentals, which also has a quantum deformed moduli space with a Pfaffian
constraint~\cite{ken}. The moduli space for supersymmetric QCD with $N=F=2$ can
be thought of as the complexification of $SU(4)/Sp(4)$, which is the
complexification of $S^5$. Similarly, the moduli space for the $Sp(2n-2)$
theories with $2n$ fundamentasl is the complexification of $SU(2n)/Sp(2n)$,
which has a deformation retract onto $SU(2n)/Sp(2n)$. It is known~\cite{witten}
that  $\pi_2\left(SU(2n)/Sp(2n)\right)=\pi_3\left(SU(2n)/Sp(2n)\right)=0$, but 
$\pi_5\left(SU(2n)/Sp(2n)\right)=\z$, so the theory has a Wess-Zumino term but
no skyrmions.

The generalization to other quantum modified theories is more involved. In
non-supersymmetric theories with a flavor symmetry $G$ broken to a subgroup
$H$, the manifold of Goldstone bosons fields is the compact manifold $G/H$. In
this case, it has been proved in general that one can always construct a
Wess-Zumino term that reproduces the correct anomalous Ward
identities~\cite{nelson}. In supersymmetric theories, there are  (non-compact)
flat directions in addition to the usual $G/H$ Goldstone boson directions. In
general, the moduli space $\CM$ is not a homogeneous space, and the unbroken
flavor group can be different at different points of $\CM$, as happens in
supersymmetric QCD with $N=F>2$. The Higgs mechanism can be used in this case
to flow to supersymmetric QCD with $N=F=2$, for which the Wess-Zumino term
exists. This indicates that a Wess-Zumino term should also exist for $N=F>2$,
but it would be useful to have an explicit construction.

\acknowledgments

I would like to thank E.~Poppitz and W.~Skiba for helpful discussions. I would
particularly like to thank K.~Intriligator for comments that simplified the
arguments in the paper. This work was supported in part by Department of Energy
grant DOE-FG03-97ER40546.

\end{document}